\documentclass[epj]{svjour}
\usepackage{graphics}
\begin{document}
\title{Energy and System-size Dependence of~Two-particle Azimuthal Correlations
  of~High-$p_{T}$ Charged Hadrons at~the~CERN~SPS}
\author{Marek Szuba\thanks{\emph{Marek.Szuba@if.pw.edu.pl}} for~the~NA49 Collaboration}
\institute{Faculty of Physics, Warsaw University of Technology, Koszykowa 75, 00-662 Warszawa, Poland}
\date{Received: date / Revised version: date}
%
\abstract{
Angular correlations of high-$p_T$ hadrons can serve as a probe of interactions of partons
with the dense medium produced in high-energy heavy-ion collisions but other sources of such correlations
exist which can be non-negligible SPS energies. In an attempt to determine the primary source, NA49
has performed an energy and system-size scan of two-particle azimuthal correlations in central $Pb+Pb$,
$Si+Si$ and $p+p$ collisions at 158$A$~GeV, as well as central $Pb+Pb$ collisions at 20$A$, 30$A$, 40$A$
and 80$A$~GeV. Moreover, results at 158$A$~GeV have been compared to UrQMD simulations.
\PACS{
      {25.75.Bh}{Hard scattering in relativistic heavy ion collisions}	\and
      {25.75.Gz}{Particle correlations and fluctuations}	\and
      {25.75.Nq}{Quark deconfinement, quark-gluon plasma production and phase transitions in relativistic heavy-ion collisions}
     } 
} 
\maketitle
\section{Introduction}
\label{sec:introduction}

It is believed that one of the signatures of a hot, dense medium expected to appear
in most central high-energy collisions of heavy ions, will be modification of properties
of jets --- highly collimated showers of particles, originating from hard scattering of partons
and therefore produced early in a collision --- as a result of their interaction with that 
medium~\cite{WangGyulassy}. Unfortunately, direct reconstruction of jets in the presence
of a large background typical of such events is an extremely challenging task. One possible answer
to this problem is studying two-particle azimuthal correlations, which allow one to extract
the jet signal from the soft background by taking advantage of the fact jet particles are strongly
correlated in azimuth. In recent years RHIC experiments have made this approach highly successful
by not only observing the expected signatures, but also providing, through further investigations,
evidence which has led to unexpected conclusions regarding the properties of the hot medium
(the ``perfect liquid'' description; see \textit{e.g.}~\cite{STARresults}, \cite{PHENIXresults}).
However, until recently (\cite{CERESfirst}, \cite{SzubaQM2008}) no analyses of this sort were performed
in the energy range of the CERN SPS.

What makes investigating hard processes challenging, even with the aid of correlations, in the energy
range of the SPS is that their contribution to particle production there is very low, \textit{i.e.}
only about 2~percent. Moreover, as SPS-energy collisions do not produce really high-$p_{T}$ particles
it is necessary to perform such analyses in the intermediate region just above 2~GeV/c in which
the share of particles originating from soft processes is still non-negligible. Finally, operating
in this transverse-momentum region in nucleus-nucleus collisions makes the analysis sensitive
to Cronin enhancement, an effect of undetermined origin and thus unknown magnitude. All of the above,
together with high uncertainties of pQCD calculations for such low momenta, make it unclear how
much of the correlation signal will come from jets and how much from other sources, such as resonances,
global momentum conservation etc.

\section{Experimental Setup}
\label{sec:experimentalSetup}

The NA49 detector is a large-acceptance spectrometer dedicated to the study of hadron production
in fixed-target nucleon-nucleon, nucleon-nucleus and nucleus-nucleus collisions at a wide range
of energies offered by the CERN SPS. During its eight years of running it has been used to register
Pb+Pb collisions at 20$A$, 30$A$, 40$A$, 80$A$ and 158$A$ GeV, $C+C$ and $Si+Si$ collisions at 40$A$
and 158$A$ GeV, along with $p+p$ and $p+Pb$ interactions at 158~GeV.

The main components of the detector are four time projection chambers, used for tracking as well
as particle identification by $dE/dx$. Two of the chambers, the so-called Vertex TPCs, are located inside
two superconducting magnets (with field intensity of 1.5 and 1.1~T, respectively) positioned along
the beam axis right downstream of the target, with the other two (``Main'' TPCs) placed further
downstream on both sides of the beam line. Two Time of Flight walls located beyond the MTPCs complement
particle identification in the intersection region of Bethe-Bloch $dE/dx$ bands of different particle species.
Centrality selection in nucleus-nucleus collisions is based on the projectile spectator energy deposited
in the Veto Calorimeter, located at the downstream end of the experiment.

The present analysis is based on the following data sets recorded by NA49:
\begin{itemize}
  \item $3\times10^6$ $Pb+Pb$ collisions, $\sigma/\sigma_{geom}$ = 0--23.5~\%, at 158$A$~GeV, from the year 2000;
  \item $265\times10^3$ $Pb+Pb$ collisions, $\sigma/\sigma_{geom}$ = 0--7.2~\%, at 80$A$~GeV, from the year 2000;
  \item $530\times10^3$ $Pb+Pb$ collisions, $\sigma/\sigma_{geom}$ = 0--7.2~\%, at 40$A$~GeV, from the year 1999;
  \item $447\times10^3$ $Pb+Pb$ collisions, $\sigma/\sigma_{geom}$ = 0--7.2~\%, at 30$A$~GeV, from the year 2002;
  \item $369\times10^3$ $Pb+Pb$ collisions, $\sigma/\sigma_{geom}$ = 0--7.2~\%, at 20$A$~GeV, from the year 2002;
  \item $220\times10^3$ $Si+Si$ collisions, $\sigma/\sigma_{geom}$ = 0--12~\%, at 158$A$~GeV, from the year 1998;
  \item $1.7\times10^6$ $p+p$ collisions ($\approx$90~\% inelastic) at~158~GeV, from the year 2000.
\end{itemize}

\section{The Method}
\label{sec:theMethod}

In all data sets except $p+p$, which is taken in its entirety, the two-particle azimuthal correlation
function is calculated for the event centrality range of 0--5~\%. The transverse momentum bins are
$2.5~GeV/c \le p_T^{trg} \le 4.0~GeV/c$ for trigger particles and $1.0~GeV/c \le p_T^{asc} \le 2.5~GeV/c$
for associates.

Following the prescription of the PHENIX Collaboration~\cite{phenixMethod} we define the correlation
function as the ratio of two distributions of $\Delta\phi = \phi_{asc} - \phi_{trg}$, where $\phi$ is
the azimuthal angle: $N_{corr}(\Delta\phi)$, in which both particles come from the same event,
and $N_{mix}(\Delta\phi)$, in which the trigger and the associate originate from different events.
Division by the mixed-event spectrum accounts for non-uniform detector acceptance. Before the ratio is calculated
the two distributions are normalised to unity; this is included in the formula in the form of appropriate
integrals.
\begin{equation}
  C_{2}(\Delta\phi) = \frac{N_{corr}(\Delta\phi)}{N_{mix}(\Delta\phi)}
    \frac{\int{N_{mix}(\Delta\phi')}\mathrm{d}(\Delta\phi')}{\int{N_{corr}(\Delta\phi')}\mathrm{d}(\Delta\phi')}
  \label{eqn:c2phi}
\end{equation}
In this analysis the mixing is accomplished with a sliding window of up to 50 most recent events in the selected
centrality bin.

By default the two distributions, and by extension the correlation function, are presented in the full
$\Delta\phi$ range of $\left[ -\frac{\pi}{2},~\frac{3}{2}\pi \right)$. On the other hand, as the correlations
under consideration are expected to be symmetric around the back-to-back axis it is possible to reduce
statistical errors by folding the function in half\footnote{Systematic uncertainties introduced
by this procedure are being investigated and will be discussed at a later date.}. In order to avoid
potential normalisation problems the folding is performed on $N_{corr}(\Delta\phi)$ and $N_{mix}(\Delta\phi)$
before they are used in Equation~\ref{eqn:c2phi}; binning of the two distributions was selected to enable
an exact bin-by-bin fold from $\left[ -\frac{\pi}{2},~\frac{3}{2}\pi \right)$ to $\left[ 0,~\pi \right]$.

All results presented here contain statistical errors only. Investigation of systematic uncertainties
is ongoing.

\paragraph{Simulated reference data} for the analysis were produced using the string-hadronic model
UrQMD~\cite{urqmd1}~\cite{urqmd2}. Version 2.3 of the model was used, which among other improvements
is the first version of the model to incorporate jet production from PYTHIA~\cite{pythia}. Four data
sets of 100,000 events each were produced:
\begin{itemize}
  \item $Pb+Pb$ collisions at 158$A$~GeV, $b = 0~fm$, jet production enabled;
  \item $Pb+Pb$ collisions at 158$A$~GeV, $b = 0~fm$, jet production disabled;
  \item $p+p$ collisions at 158~GeV, $b = 0~fm$, jet production enabled;
  \item $p+p$ collisions at 158~GeV, $b = 0~fm$, jet production disabled;
\end{itemize}

\section{Results}
\label{sec:results}

\subsection{System-size Scan}
\label{sec:systemSizeScan}

\begin{figure*}
  \resizebox{\textwidth}{!}{
    \includegraphics{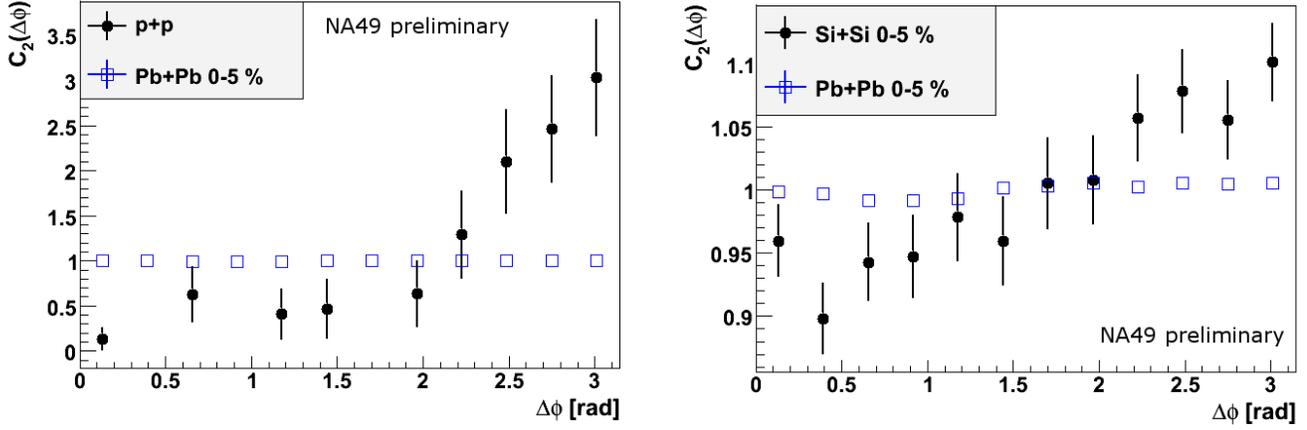}
  }
  \caption{Two-particle correlation functions from $p+p$ (left) and central $Si+Si$ (right) events
    at 158$A$~GeV, compared to central-$Pb+Pb$ results at the same energy.}
  \label{fig:systemSizeScan}
\end{figure*}

Figure~\ref{fig:systemSizeScan} shows two-particle azimuthal correlation functions from $p+p$
and central $Si+Si$ collisions at 158$A$~GeV, compared to those obtained earlier from central lead--lead
data at the same energy~\cite{SzubaQM2008}. Overall strength of the correlation becomes significantly larger
as the system size decreases. Moreover, no flattening of the away-side peak, as present in central
heavy-ion events, is visible in $Si+Si$ and $p+p$ collisions --- indeed, the peak becomes narrower
with decreasing system size.

\subsection{Energy Scan}
\label{sec:energyScan}

\begin{figure*}
  \resizebox{\textwidth}{!}{
    \includegraphics{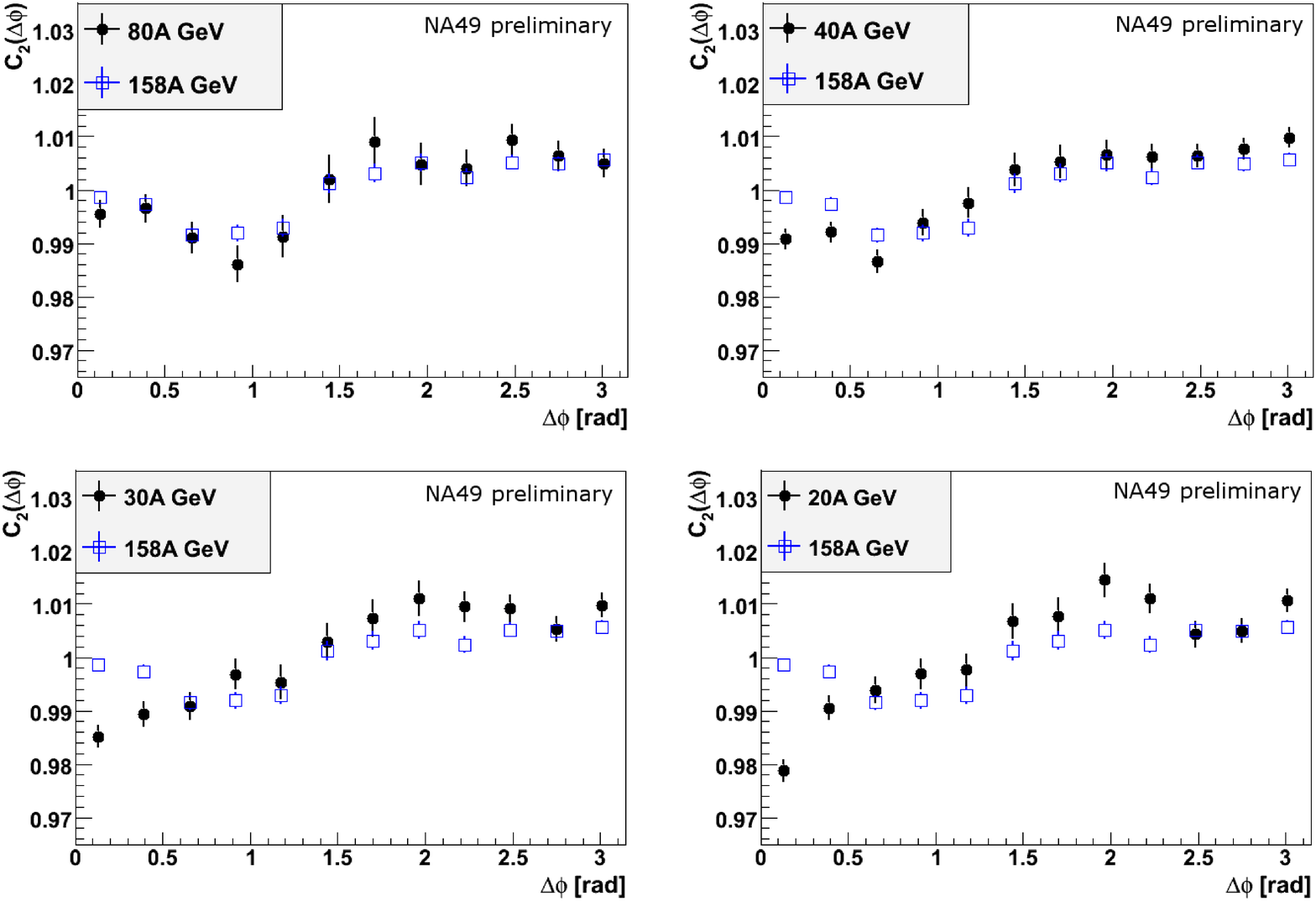}
  }
  \caption{Two-particle correlation functions from central $Pb+Pb$ events at 80$A$ (top left), 40$A$ (top right),
  30$A$ (bottom left) and 20$A$ (bottom right) GeV, compared to results from the same system at 158$A$~GeV.}
  \label{fig:energyScan}
\end{figure*}

In Figure~\ref{fig:energyScan} the correlation function from central $Pb+Pb$ collisions at 158$A$~GeV is
compared to results from the same system at 80$A$, 40$A$, 30$A$ and 20$A$~GeV. The near-side peak appears
to turn into depletion with decreasing energy, then again the shape and amplitude of away-side enhancement
remains mostly unchanged throughout the scan. Should the flattening of the latter be considered a signature
of a hot, dense medium, these results are at odds with present-day expectations that such a medium is produced
only at higher energies. On the other hand, the observed away-side enhancement appears to be consistent with
global momentum conservation.

\subsection{Comparison with UrQMD}
\label{sec:comparisonWithUrQMD}

\begin{figure}
  \resizebox{0.5\textwidth}{!}{
    \includegraphics{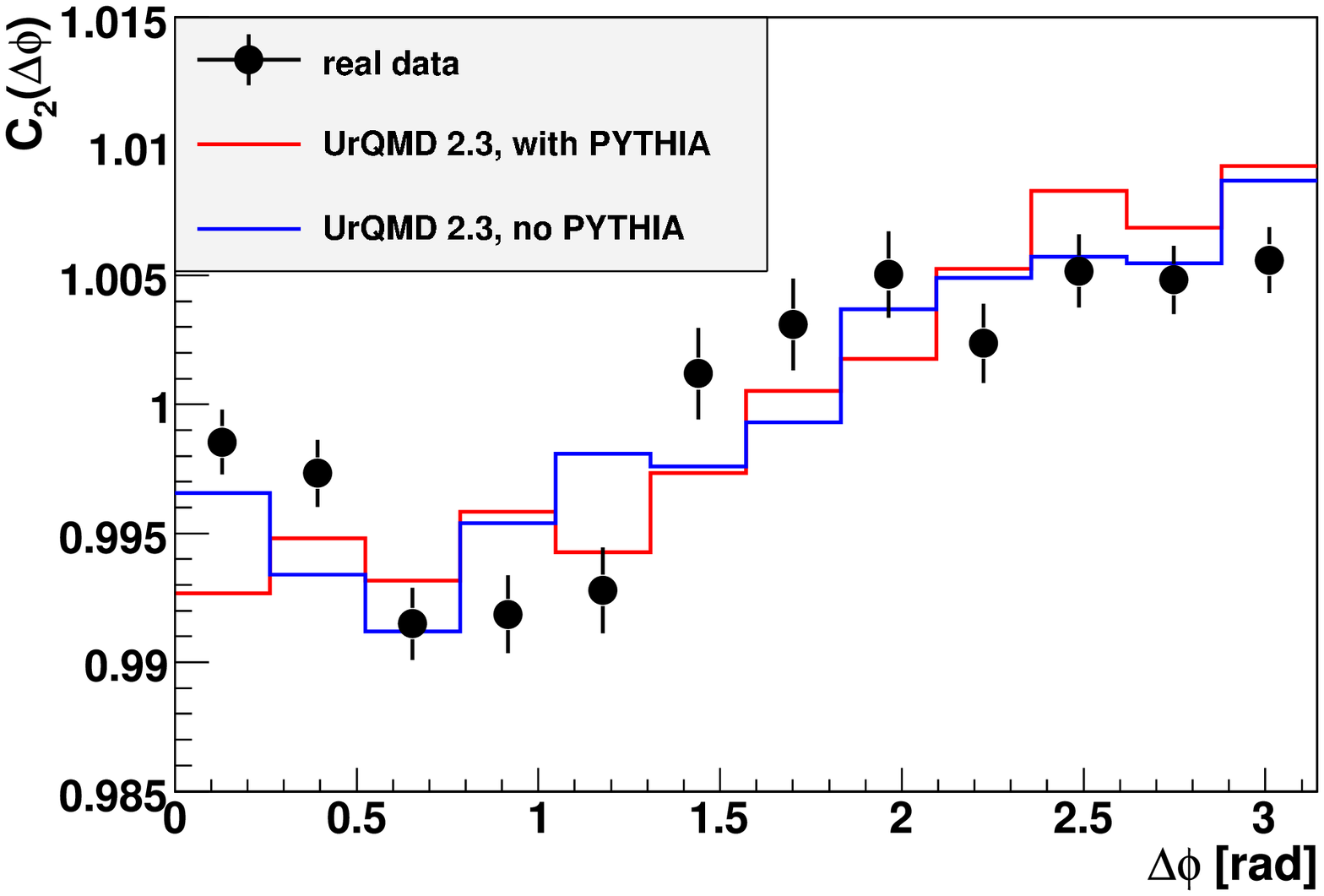}
  }
  \caption{Comparison of experimental and simulated correlation functions for central lead--lead collisions
    at 158$A$~GeV. Black points: experimental data, red line: UrQMD with jets, blue line: UrQMD without jets.}
  \label{fig:urqmd_Pb}
\end{figure}

\begin{figure}
  \resizebox{0.5\textwidth}{!}{
    \includegraphics{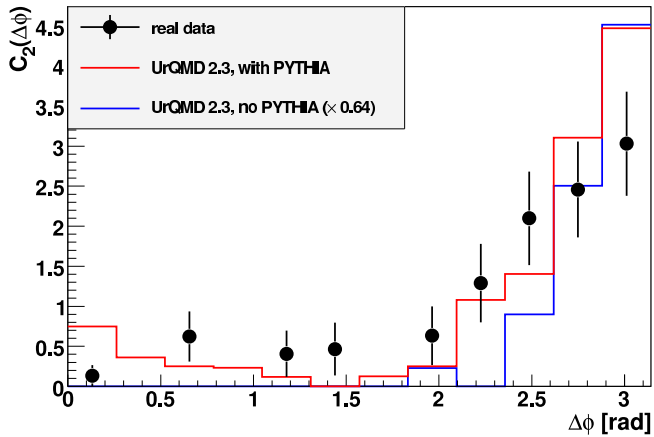}
  }
  \caption{Comparison of experimental and simulated correlation functions for proton--proton collisions
    at 158~GeV. Black points: experimental data, red line: UrQMD with jets, blue line: UrQMD without jets
    (scaled to match the amplitude of the away-side peak of with-jets data).}
  \label{fig:urqmd_p}
\end{figure}

A comparison of real-data azimuthal correlation functions from central-$Pb+Pb$ and $p+p$ collisions with results
obtained from UrQMD simulations can be found in Figures~\ref{fig:urqmd_Pb} and~\ref{fig:urqmd_p},
respectively. For both systems fairly good agreement can be observed between the data and the simulations,
regardless of whether production of jets was enabled in the model or not. These results are again
consistent with momentum conservation, however due to the nature of UrQMD it is not possible to say at this
point whether it is global or local.

\section{Summary}
\label{sec:summary}

A system-size and energy scan of two-particle correlation functions from NA49 has been performed.
It was found that flattening of the function's away-side peak can be observed in central
heavy-ion collisions even at low SPS energies, where no hot, dense medium is expected. On the other
hand, all results appear to be consistent with global momentum conservation.

Comparison with results from UrQMD show good agreement between the model and the experiment. Furthermore,
the former exhibit little dependence on presence or absence of jets in the sample, favouring the
momentum-conservation hypothesis. Further analysis involving statistical models is required in order
to determine whether this conservation is global or local.

\end{document}